# Measurement of the neutron lifetime using a gravitational trap and a low-temperature Fomblin coating

A. Serebrov[1], V. Varlamov[1], A. Kharitonov[1], A. Fomin[1], Yu. Pokotilovski[2], P. Geltenbort[3], J. Butterworth[3], I. Krasnoschekova[1], M. Lasakov[1], R. Tal'daev[1], A. Vassiljev[1], O. Zherebtsov[1]

[1] *Petersburg Nuclear Physics Institute, Russian Academy of Sciences, Gatchina, Leningrad District, 188300, Russia*
[2] *Joint Institute for Nuclear Research, Dubna, Moscow Region, 141980, Russia*
[3] *Institut Max von Laue – Paul Langevin, B.P.156, 38042 Grenoble Cedex 9, France*

**Abstract**

We present a new value for the neutron lifetime of 878.5 ± 0.7 $_{stat.}$±0.3 $_{syst}$. This result differs from the world average value (885.7 ± 0.8 s) by 6.5 standard deviations and by 5.6 standard deviations from the previous most precise result [1]. However, this new value for the neutron lifetime together with a β-asymmetry in neutron decay, $A_0$, of -0.1189(7) [2] is in a good agreement with the Standard Model.

## 1. Introduction

The decay of the free neutron into a proton, an electron and an antineutrino is related to the weak interaction process. In the Standard Model the probability of this process or the lifetime of the free neutron is related to the vector $G_V$ and axial $G_A$ weak interaction coupling constants. The neutron lifetime has important implications in particle physics, in neutrino-induced reactions and in cosmology. The neutron lifetime together with angular correlation coefficients of the decay of a polarised neutron allows deduction of the axial and vector coupling constants only from neutron decay data. The main element of the Kobayashi-Maskawa matrix, $V_{ud}$, has to be determined with the highest accuracy to check for eventual deviations from the Standard Model which are currently under discussion [3].

The observed deviation from the unitarity condition for the Cabibbo-Kobayashi-Maskawa matrix using present data is about 2.7 standard deviations. The origin of this deviation is unclear. This situation requires more precise measurements of $\beta$-asymmetry - $A_0$ and new measurements of the neutron lifetime.

Today, the weighed mean value of the neutron lifetime is 885.7(8) s. The accuracy of this world average was improved by the lifetime experiment of a group from KIAE, Russia [1]. Their result (885.4 ± 0.9$_{stat.}$ ± 0,4$_{syst}$) has an accuracy that is at least 3 times better than that of the other contributing experiments. So the present world mean value of the neutron lifetime is mainly determined by the result of only one experiment, therefore the new experimental measurements are important.

## 2. Experimental set-up

The present measurements were carried out at the high flux reactor at ILL in Grenoble, France using the PF2/MAM instrument; the experimental set-up is sketched in Fig. 1. It is a gravitational trap for UCN and at the same time it can be used as a differential gravitational spectrometer. Therefore the distinguishing feature of this experiment is the ability to measure the UCN energy spectrum after its storage in the trap.

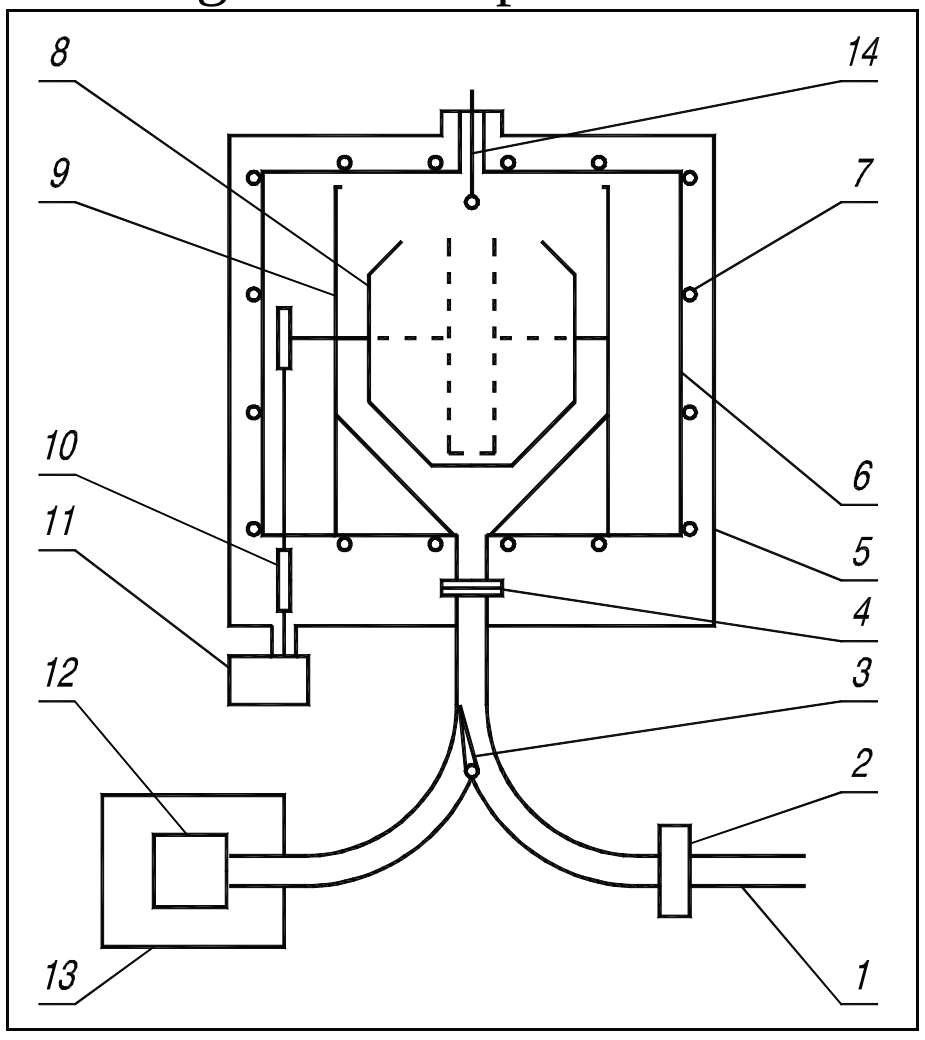

Fig. 1. The Scheme of “Gravitrap”, the gravitational UCN storage system. 1: neutron guide from UCN Turbine; 2: UCN inlet valve; 3: beam distribution flap valve (shown in the filling position); 4: connection unit; 5: “high” vacuum volume; 6: “rough” vacuum volume; 7: cooling coils; 8: UCN storage trap (the narrow cylindrical trap is shown by a dashed line); 9: cryostat; 10: mechanics for trap rotation; 11: stepping motor; 12: UCN detector; 13: detector shielding; 14: evaporator.

The UCN storage trap *8* is mounted inside a cryostat vacuum vessel *9*. The trap *8* has a window that can be rotated about a horizontal axis so that UCN are held in the trap by gravity when the trap window is in its upper position.

UCNs enter the trap via the neutron guide *1*, the opened UCN inlet valve *2* and the distribution flap valve *3*. Filling takes place when the trap window is in the down position. After the trap is filled it is rotated into the up position.

A double walled vacuum system was used with separate "high" *6* and "rough" *5* vacuum vessels. The pressure in the cryostat vacuum vessel was $5{\cdot}10^{-6}$ mbar; at this pressure, the residual gas has a small effect (0.4 s, see below) on storage time for the UCN in the trap. To cool the trap we used heat exchange between the trap and the cryostat tank; to do this helium gas was flowed through the cryostat vacuum vessel and removed before carrying out the neutron lifetime measurements.

The height of the trap window relative to the trap bottom defines the maximum energy of UCN that can be held in the trap. Different window heights correspond to different cut-off energies for the UCN spectrum. Such a rotatable trap is a gravitational spectrometer. The spectral dependence of the storage time can be measured by turning the trap window downward in steps. The trap was kept in each intermediate position during 100-150 s to detect UCN in the corresponding energy range. The same procedure also measures the spectrum of the trapped UCN.

The neutron lifetime is measured with the size extrapolation method using two sizes of UCN trap. The first is a quasi-spherical trap consisting of a cylinder about 84 cm in diameter and 26 cm wide, capped by two truncated cones each 22 cm high, with small diameters of 42 cm and the second a 76 cm diameter cylindrical trap that was 14 cm long between its end faces. The second trap increases the neutron collision rate with the walls of the trap by a factor of about 2.5. The narrow cylindrical trap is shown in Fig. 1 by a dashed line.

A typical count rate diagram during the UCN storage cycle is shown in Fig. 2. First the trap is filled with UCN with the hole in the down position. Then the trap is rotated to the monitoring position where the height of the trap window is 10 cm lower than when in the holding position with the hole upward. The filling process can be observed by means of the detector *12* through the slits in the distribution valve. When the trap reaches the monitor height the distributive valve is changed to the detection position. The trap is kept in the monitoring position for 300 s. During this period the neutrons whose energy exceeds the gravitational barrier of the trap escape (see Fig. 2). Then the trap is rotated to the holding position. Overall this process takes about 460 s and the counting rate is shown on a logarithmic scale on the left side of Fig. 2; the counting rate for the subsequent procedures (700-3160 s) is shown on a linear scale on the right side, in order to show more details. After a short (top part of Fig. 2) or long (bottom part of Fig. 2) holding time, the trap is rotated to five successive positions and is held in each position for 100 to 150 s in order to count UCN. The neutrons counted after each rotation have a different average energy. When the UCN trap is empty the background measurement is started. The angle positions of the trap are: $\theta = 30^o$, (monitoring position), $\theta = 40^o$, $E_{UCN}$ = 58 cm; $\theta = 50^o$, $E_{UCN}$ = 52 cm; $\theta = 60^o$, $E_{UCN}$ = 46 cm; $\theta = 75^o$, $E_{UCN}$ = 39 cm; $\theta = 180^o$, $E_{UCN}$ = 25 cm. The angles were chosen so as to obtain similar counts for each portion of the UCN spectrum (unfortunately, the third portion of was not successfully optimised as may be seen in Fig. 2).

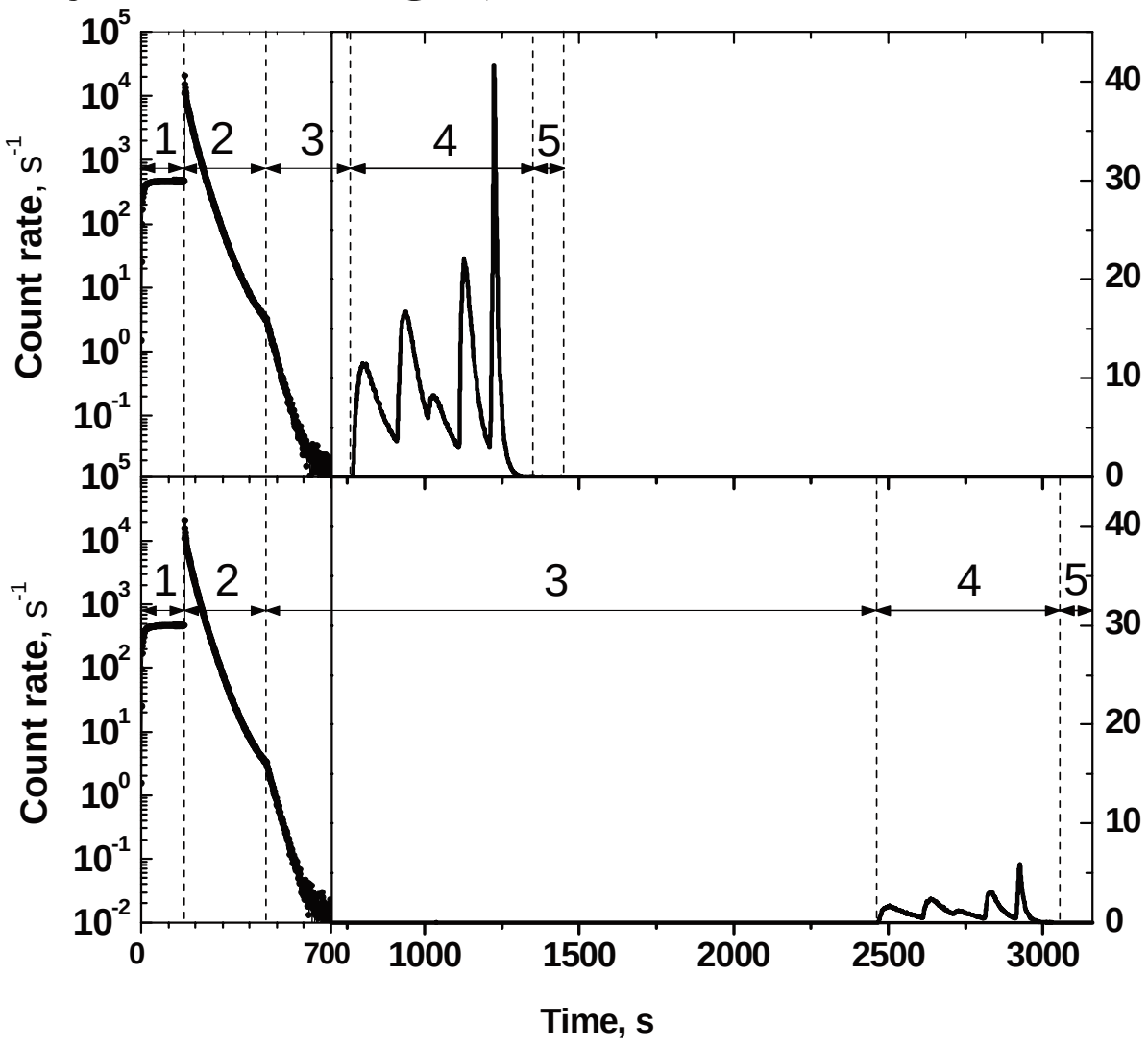

Fig. 2. Time diagrams of the storage cycle for two different holding times. 1: filling 160 s (time of trap rotation (35 s) to monitoring position is included); 2: monitoring 300 s; 3: holding 300 s or 2000 s (time of trap rotation (7 s) to holding position is included); 4: emptying has 5 periods 150 s, 100 s, 100 s, 100 s, 150 s (time of trap rotation (2.3 s, 2.3 s, 2.3 s, 3.5 s, 24.5 s) to each position is included); 5: measurement of background 100 s.

## 3. Methods of the neutron lifetime extrapolation

### 3.1 Basic relationships

Any neutron lifetime measurement using the UCN storage method is based on the rather simple equation (1). The total probability of UCN losses $\tau_{st}^{-1}$ is broken up into two parts: the first part is the probability of neutron beta-decay $\tau_n^{-1}$ and the second one is the probability of other possible UCN losses $\tau_{loss}^{-1}$:

$$\tau_{st}^{-1} = \tau_n^{-1} + \tau_{loss}^{-1} \qquad (1)$$

The UCN storage time $\tau_{st}$ is calculated from the measurements of the number of neutrons remaining in the trap after different holding times:

$$\tau_{st} = (t_2 - t_1)/ln(N_1/N_2) \ , \qquad (2)$$

where $N_1$ and $N_2$ are the numbers of neutrons remaining in the trap after holding times $t_1$ and $t_2$ respectively. There is no need to know the efficiency of the UCN detector, the UCN loss rate during its transport to the detector, etc., since equation (2) uses only the ratio of $N$.

As the UCN are held by the trap material then $\tau_{loss}^{-1}$ contains the probability of losses at the trap walls:

$$\tau_{loss}^{-1} = \mu(T,E) \cdot \nu(E) , \qquad (3)$$

where $\mu(T,E)$ is the UCN loss factor per reflection, which depends on the UCN energy and the temperature of walls $T$, and $\nu$ is the UCN collision frequency, which depends on the UCN energy and the size of the trap. The UCN loss factor for one reflection can be presented in the following well known form [4], which was derived on the basis that the reflection of UCN takes place from a potential step with real ($U_0$) and imaginary ($W$) parts

$$\mu(y) = \frac{2\eta}{y^2} \cdot (\arcsin y - y\sqrt{1-y^2}) =$$

$$= \begin{cases} \pi\eta, \ (y \to 1) \\ \approx \frac{4}{3}\eta y, \ (y \leq 1) \end{cases} \qquad (4)$$

where $\eta$ is the ratio of the imaginary and real parts of the potential or scattering amplitude and

$$\eta = \frac{W}{V} = \frac{b'}{b}, \ y = \left(\frac{E}{U_0}\right)^{\frac{1}{2}}.$$

The UCN loss factor in equation (4) is averaged over the angles of incidence. Using the optical theorem, the imaginary part of the scattering amplitude can be written in the following form [4]: $b' = \frac{\sigma_{abs} + \sigma_{upscat}(T)}{2\lambda}$. The absorption and up-scattering cross sections are proportional to the neutron wavelength ($\lambda$), therefore $b'$ and $\eta$ do not depend on $\lambda$ or neutron energy E. But $\eta$ is a function of temperature ($\eta = \eta(T)$) because of the temperature dependence of the up-scattering cross section ($\sigma_{upscat}(T)$).

We can now rewrite equation (3) as the product of separate energy and temperature dependent factors:

$$\tau_{loss}^{-1} = \eta(T) \cdot \gamma(E), \qquad (5)$$

where $\eta(T)$ is written for the energy independent part of the loss factor and $\gamma(E)$ for the effective frequency of collisions, which depends on the UCN energy and the trap size. The neutron lifetime value can be obtained by a linear extrapolation of $\tau_{st}^{-1}$ to zero value of $\gamma$. Different values of the UCN effective collision frequency $\gamma$ are obtained using traps of different size and using different UCN energies. The UCN loss factor $\eta$ is then the slope of the extrapolation line. The UCN effective collision frequency $\gamma$ can be calculated.

### 3.2 Method of energy extrapolation

To calculate the energy dependence of the UCN losses, $\tau_{loss}^{-1}$, we used numerical and Monte Carlo methods taking into account the motion of UCN in the gravitational field. The simple relation $\tau_{loss}^{-1} = \mu(E) \cdot \nu(E)$ was used to calculate the probability of losses. The frequency of collisions $\nu(E)$ with the element of surface $dS$ is related to the flux directed at the surface, $\frac{1}{4}\mathrm{v}\rho(\mathrm{v})dS$, where $\rho(\mathrm{v})$ is the UCN density and dependent on the UCN velocity. In the gravitational field the distribution of UCN density is proportional to $\sqrt{\frac{E-mgh}{E}}$, where $E$ is the UCN energy at the bottom of the trap and $h$ is the height measured from the bottom of the trap [4]. This equation has to be integrated and normalised as UCN have different kinetic energies at different $h$:

$$\tau_{loss}^{-1}(E) = \frac{\int_0^E \mu(E-h') \cdot \mathrm{v}(E-h') \cdot \rho(E-h') dS(h)}{4\int_0^E \rho(E-h') dV(h)}, \qquad (6)$$

where $h' = mgh$.

For comparison with experimental data we have to integrate equation (6) over the energy interval of each measurement and to take into account the real spectral distribution of UCN density in the trap. The UCN spectrum in the trap was measured just after the monitoring process by means of the procedure shown in the top part of Fig. 2. The modification of the spectrum during the process of UCN storage in the trap was taken into account and gives a small correction to the calculated function $\gamma$ ($\frac{\Delta\gamma}{\gamma} = 0.1\%$ or 0.01 s in the extrapolated neutron lifetime). Additional corrections to the calculated function $\gamma$ were made, to take into account incomplete emptying of UCN from each energy interval, as can be seen in Fig. 2. It can bring the correction in the extrapolated neutron lifetime up to 0.7 s, but using the detailed Monte Carlo calculations and method of size extrapolation (see below) we can reduce this correction more than order of magnitude and we will have the uncertainty of extrapolation to neutron lifetime 0.24 s.

Using the calculated values of losses for different energies, $\tau_{loss}^{-1} = \eta(T)\cdot\gamma(E)$, we can extrapolate the measured data to give the neutron lifetime. The result from the extrapolation depends on the function $\mu(E)$, which may be a little bit different in reality from the form assumed, for example $\mu'(E)$. To reduce the systematic effect that could arise from the energy dependence $\mu'(E)$ we have to consider how to exclude the energy dependence from the extrapolation.

### 3.3 Method of size extrapolation

To exclude the energy dependence, $\mu'(E)$, we can make an extrapolation to zero loss using the data for the same UCN energy for traps with different sizes. Using equations (1) and (5) for two traps with storage times $\tau_1$ and $\tau_2$:

$$\tau_1^{-1}(E) = \tau_n^{-1} + \eta\gamma_1(E) \qquad (7)$$

$$\tau_2^{-1}(E) = \tau_n^{-1} + \eta\gamma_2(E) \qquad (8)$$

we obtain:

$$\tau_n^{-1} = \tau_1^{-1}(E) - \left(\tau_2^{-1}(E) - \tau_1^{-1}(E)\right)/[\gamma_2(E)/\gamma_1(E) - 1] \qquad (9)$$

That is, we can exclude the effect of the energy dependence $\mu'(E)$ almost completely, because the final result for the neutron lifetime depends only on the ratio $\gamma_2(E)/\gamma_1(E)$. It is easy to show that in the case without gravity, using equations (7) and (8) for two different traps and for a specific defined energy allows us to completely exclude the $\mu'(E)$ function. In the real case with gravity, complete exclusion of the $\mu'(E)$ dependence is impossible because of the integral equation (6). However, the residual effect of the influence of the $\mu'(E)$ dependence on the neutron lifetime is negligibly small (0.14 s) in comparison with the statistical accuracy of the measurements (0.7 s).

The method of size extrapolation based on two traps allows large suppression of the systematic errors arising from uncertainties in our knowledge of $\mu'(E)$.

## 4. The low-temperature Fomblin coating of traps

In the experiment, we have used a new type of wall coating, a low-temperature Fomblin (LTF) that can be evaporated onto the surface in vacuum. This perfluorinated oil has a composition containing only $C$, $O$ and $F$ and thus a low neutron capture cross section. Earlier investigations [5] of several types of LTF found that quasi-elastic UCN scattering and thermal inelastic scattering are significantly lower at temperatures below -120˚C than for ordinary Fomblin oil close to room temperature. The quasi-elastic UCN scattering is suppressed completely below –120˚C [5] and the expected UCN loss coefficient $\eta$ due to up-scattering is about $2\cdot10^{-6}$ [6,7].

The new type of LTF used in our experiment has a molecular weight M = 2354 and a vapour pressure $P$ = $1.5\cdot10^{-3}$ mbar at room temperature. For preparation of the trap coating, the oil was delivered to an evaporator through a vertical tube driven by $He$ pressure. A spherical evaporator with small holes was heated up to +140ºC by means of an electric heater, and then 3 $cm^3$ of oil were ejected onto the interior trap walls held at a temperature of –150ºC. The evaporator was moved up and down during the deposition to obtain a uniform coating.

To test the oil film quality we measured the UCN lifetime in a titanium-coated $Cu$ trap with various coating of LTF. Titanium has a negative scattering length and UCN cannot be held in this trap when uncoated. The trap was a cylinder of diameter of 76 cm and length 50 cm. A stable storage time $\tau_{st.} = 869.0 \pm 0.5$ s was reached after several

LTF evaporations (total thickness of 15 μm) at a wall temperature of between -140ºC and -150ºC, then warming the trap up to room temperature and finally re-cooling to -160ºC. Due to this process, the oil at room temperature filled all the gaps and cracks in the wall and formed a perfect surface. Additionally, the LTF have been degassed in a thin layer at room temperature. The coating is very stable and no significant change of storage time was found over an 8 days observation period. Further evaporation did not change the storage time.

For the final measurements we used *Be*-coated traps (quasi-spherical and narrow cylindrical). Since *Be* is a good UCN reflector, we can use even lower wall temperatures, making the development of any micro-cracks in the coating less detrimental to the storage time. Using the *Be*-coated trap we studied the temperature dependence of the storage time for the quasi-spherical trap with an LTF coating (Fig. 3a). The LTF was deposited at –140ºC, and then the trap was slowly warmed up to –50ºC and finally cooled again to –160ºC. In this way we covered layer defects by the oil when it was sufficiently liquid. Following this temperature cycle we obtained a higher storage time, $872.2 \pm 0.3$ s, than immediately after evaporation ($850 \pm 1.8$ s). Taking into account the different sizes of the *Ti*- and *Be*-traps we find that difference of the expected storage time for *Ti*-trap and the obtained storage time is $1.9 \pm 0.6$ s. It means that the uncoated part of *Ti*-trap surface is $(4.4 \pm 1.3) \cdot 10^{-7}$ only. This indicates that reproducible oil films can be obtained due to the oil surface tension independently of substrate material and the shape of the trap. Therefore we have no reason to consider different loss factors, $\eta$, for the different traps with *Be*-substrate under LTF.

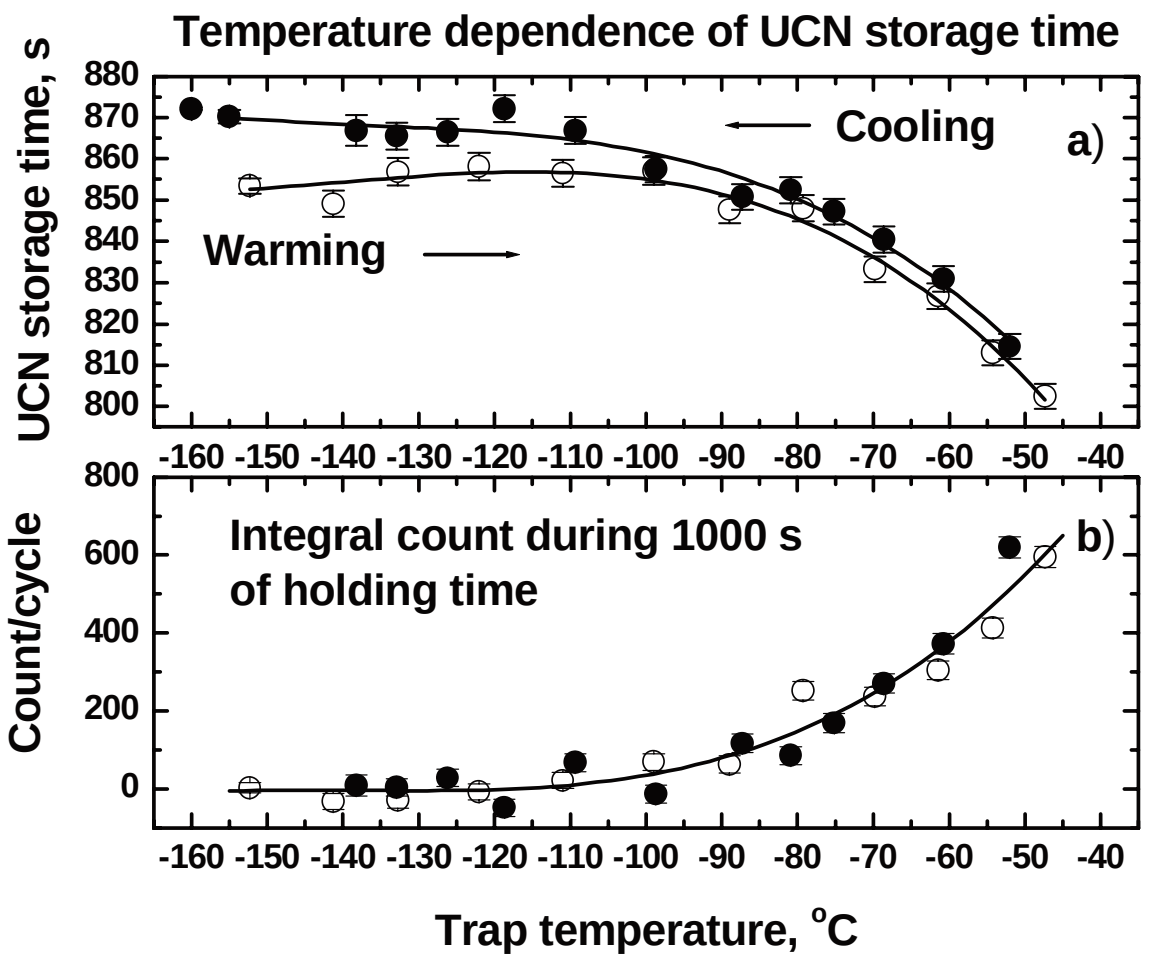

Fig. 3. a) Temperature dependence of UCN storage time during cooling and warming.
b) Temperature dependence of integral count during 1000s of holding time.

In the course of these studies with the *Be*-trap the quasi-elastic scattering by the liquid LTF was measured. Fig. 3b shows the number of UCN leaving the trap during a 1000 s holding time with the trap in the upright position as a function of trap wall temperature. During this time we observed an excess count-rate falling exponentially with the UCN storage time in the trap; this additional count rate is due to UCN suffering a small energy transfer in quasi-elastic scattering from the liquid wall. These neutrons escape from the trap and produce an increase in the detector counting rate. The process disappears at temperatures below about -120ºC. This is in qualitative agreement with the measurement of quasi-elastic UCN scattering on this oil in our previous work [5]. Quasi-elastic scattering is seen to be insignificant for T<-120ºC. In our measurements we used a temperature of -160ºC to be sure that quasi-elastic scattering would not affect our measurements. However, using this lower temperature is undesirable because of the possibility of cracks in the coating, as it loses its plastic properties.

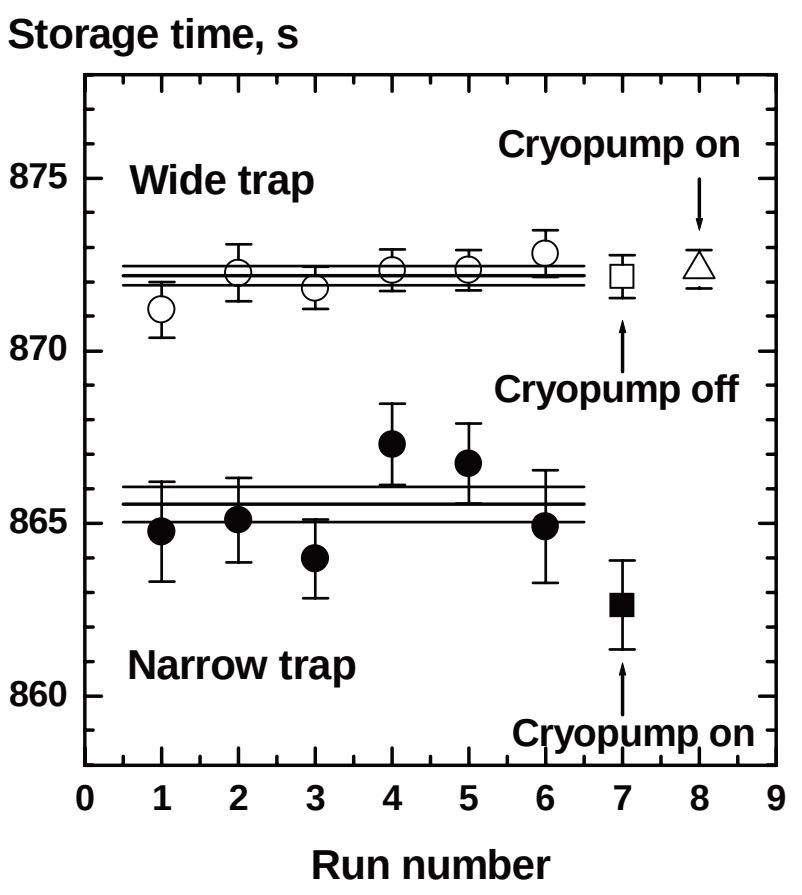

Fig. 4. Demonstration of stability of the LTF coating during the measurements. UCN storage times for wide and narrow traps plotted versus run number.

The stability and integrity of the coating of the different traps is the most important condition for validity of the size extrapolation method for the neutron lifetime measurement. Therefore the quality of the LTF coatings was verified many times during the course of the measurements. Fig. 4 shows eight storage time results for the quasi-spherical trap and seven storage time results for the narrow trap. The measurements were carried out after new evaporations, warming and cooling, new evaporations and so on. As a final improvement, the pressure in the trap was reduced from $5 \cdot 10^{-6}$ to $3 \cdot 10^{-7}$ mbar by installing a LHe cryopump near the storage volume. The storage times over the course of the neutron lifetime experiment agreed within about 1 s for the wide trap and by a little bit more for the narrow trap. This gives confidence in the stability and the reproducibility of LTF coating for the different traps.

## 5. Results of measurements and extrapolation to the neutron lifetime

The results of measurements of the UCN storage time for different energy intervals and for different traps (wide and narrow) are presented in Fig. 5 as a function of effective frequency of collisions $\gamma$. The extrapolation of all data to the neutron lifetime gives a value of $877.60 \pm 0.65$ s with a $\chi^2$ of 0.95. This means that joint extrapolation is possible. Nevertheless, we have done the energy extrapolation for each trap and on combining both results we obtain $875.55 \pm 1.6$ s.

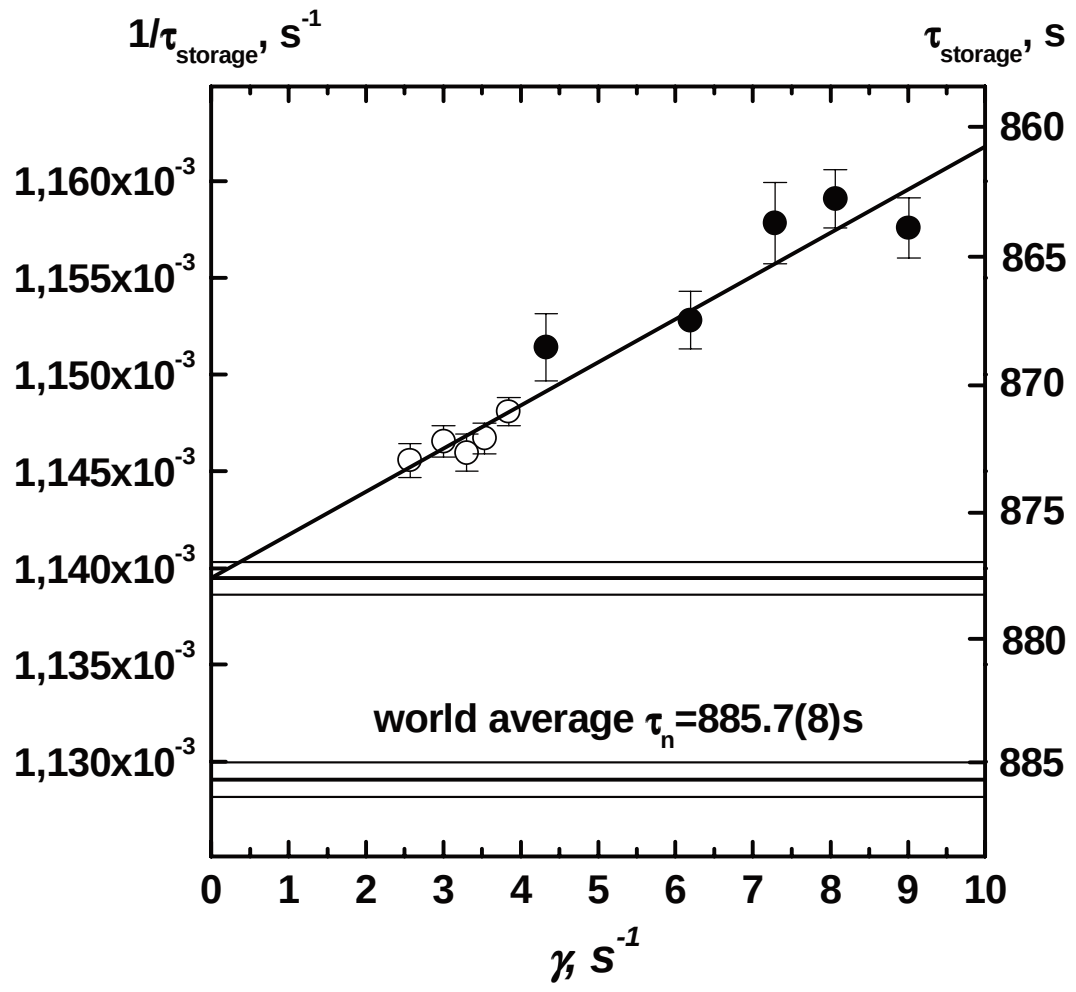

Fig. 5. Result of extrapolation to the neutron lifetime using joint energy and the size extrapolation method. Measurements made with a spherical (open circles) and cylindrical (filled circles) traps.

For the size extrapolation method we have to connect the values for the different traps for the same UCN energy interval, and then to calculate the average value of all determinations of the neutron lifetime. Fig. 6 shows the results of the size extrapolation to the neutron lifetime for the different energy intervals. The average value of the neutron lifetime from the size extrapolation method is $878.07 \pm 0.73$ s.

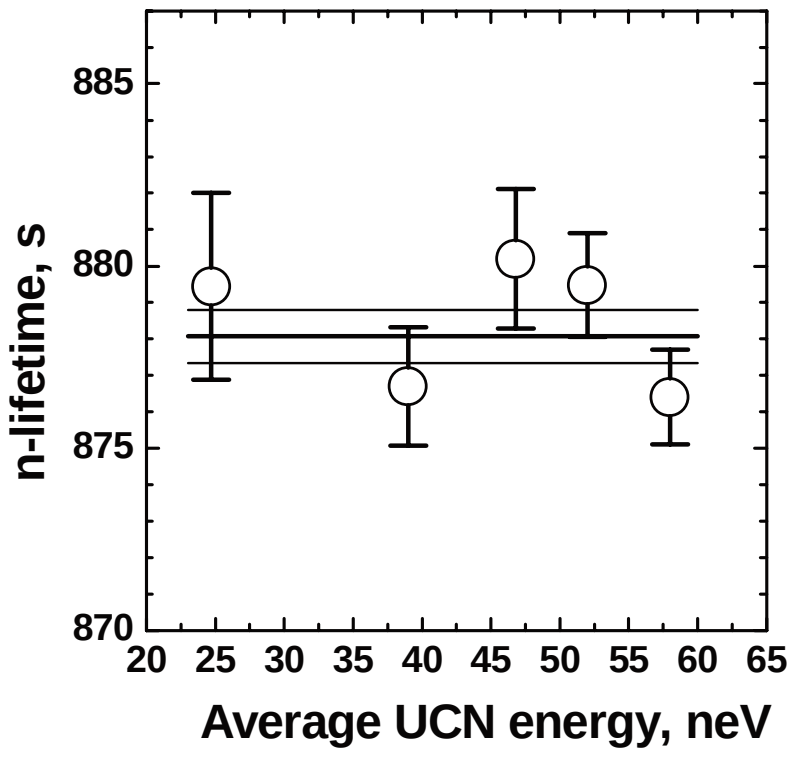

Fig. 6. Extrapolated values for the neutron lifetime for different average UCN energies using the size extrapolation method. The solid line corresponds to the final neutron lifetime fit.

The results obtained from the two methods are different by 1.5 standard deviations which can not be considered as disagreement. The loss factor, $\eta$ = $2\cdot 10^{-6}$, obtained in this experiment is in agreement with that obtained in a transmission experiment [7]. As the final value for the neutron lifetime we prefer to use the more precise result from the size extrapolation, moreover this has a rather weak dependence on $\mu(E)$ and we consider it to be the more reliable. The experimental studies of the problem of UCN losses during storage in material traps are presented in our work [8].

## 6. Monte Carlo simulation of the experiment and systematic errors

To both estimate the accuracy of and to check the reliability of the size extrapolation method using the calculated $\gamma$-function, we have used a Monte Carlo (MC) simulation of the experiment.

In the MC-simulations the behaviour of neutrons has been described taking into account the gravitational field, the form of the storage traps, losses in the trap $\eta$ = $2\cdot 10^{-6}$, the geometry of the secondary vessel and the UCN guide for transporting the UCN to the detector. As a result we can simulate directly the measurements and obtain the time diagram as shown in Fig. 2. The UCN storage times in the traps have been calculated in the same way as done in the experiment, and the extrapolation to the neutron lifetime using the calculated $\gamma$-function made as well. The single free parameter in the MC-simulation is the coefficient of diffuse scattering at the interaction with the trap surface. Knowing whether or not the probability of mirror reflection is extremely high is important; for example at 99.9 %, a highly correlated behaviour of the UCN inside the trap is possible and the performance is very difficult to predict.

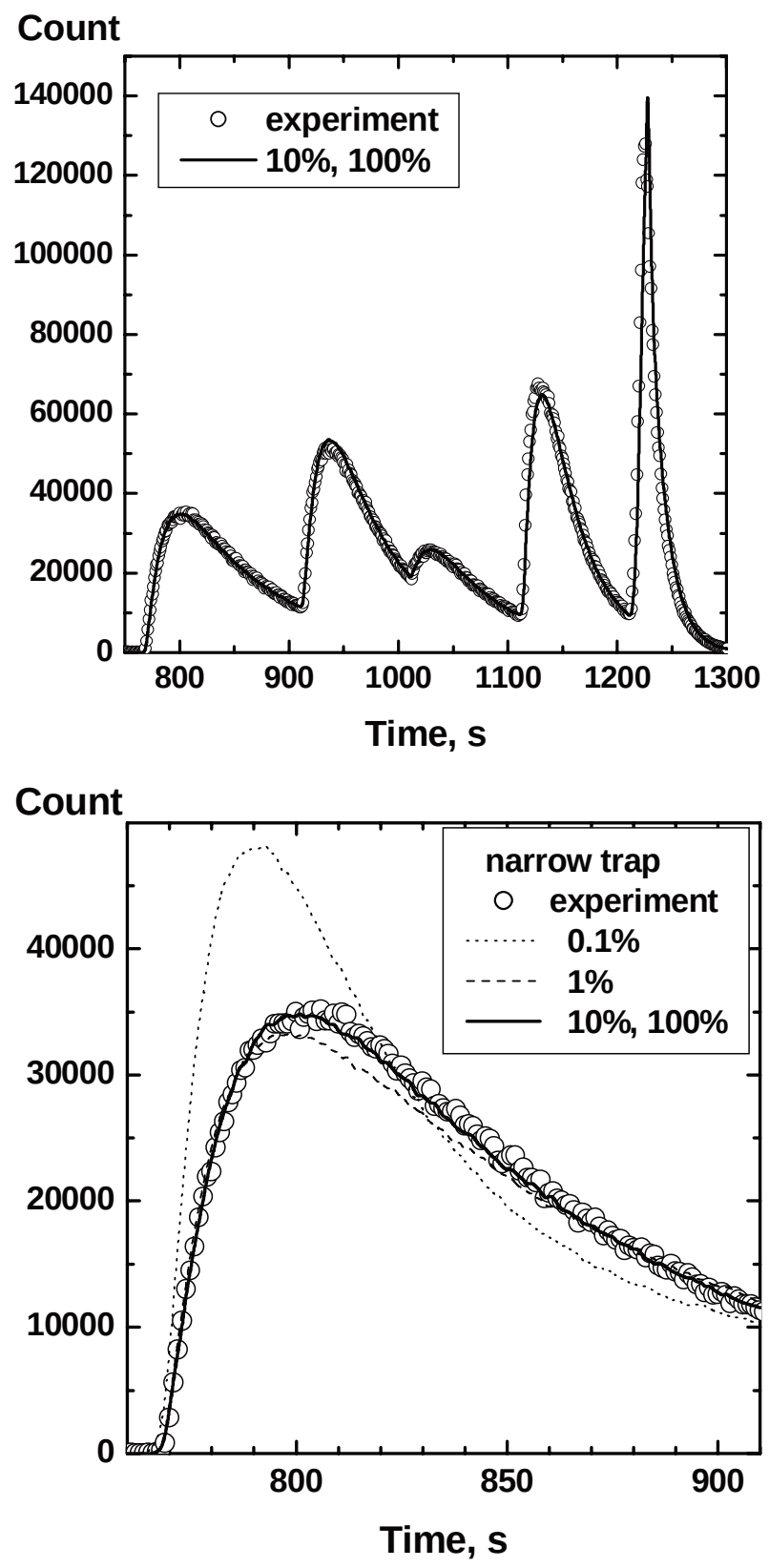

Fig. 7. Monte Carlo simulation of the experiment. Modelling the trap emptying process for the narrow cylindrical trap.

A comparison between the results of MC simulations with various values for the diffuse scattering probability and the experimental results in Fig. 2, allows us to conclude that the probability of diffuse scattering of UCN on the LTF coating is 10% or more. Fig. 7a shows the comparison of the experimental diagram and MC simulation for diffuse scattering coefficients of 10 % and 100 %; both successfully describe the experiment. However, with a probability of diffuse scattering of 0.1 %, the agreement with the experimental results is not satisfactory. The results of this calculation for the first part of the time diagram, which is the most sensitive to the effect of mirror reflections, are shown on a large scale in Fig. 7b. The final simulation of the experiment was done using probabilities of diffuse reflection of 10 % and 1 %. The simulated storage times extrapolated to the neutron lifetime for both the wide cylindrical and the narrow cylindrical traps and for the five different UCN energy intervals of the experiment are shown in Fig. 8. To simplify the MC calculations we used a cylinder rather than the quasi-spherical form for the wide trap. The final analysis of the simulated measurements reproduced the neutron lifetime value assumed in the calculation with an accuracy of $\pm 0.236$ s. This accuracy is limited by the statistical accuracy of the MC calculation. That is, the systematic uncertainty of the size extrapolation method using the calculated $\gamma$-function is $\pm 0.236$ s.

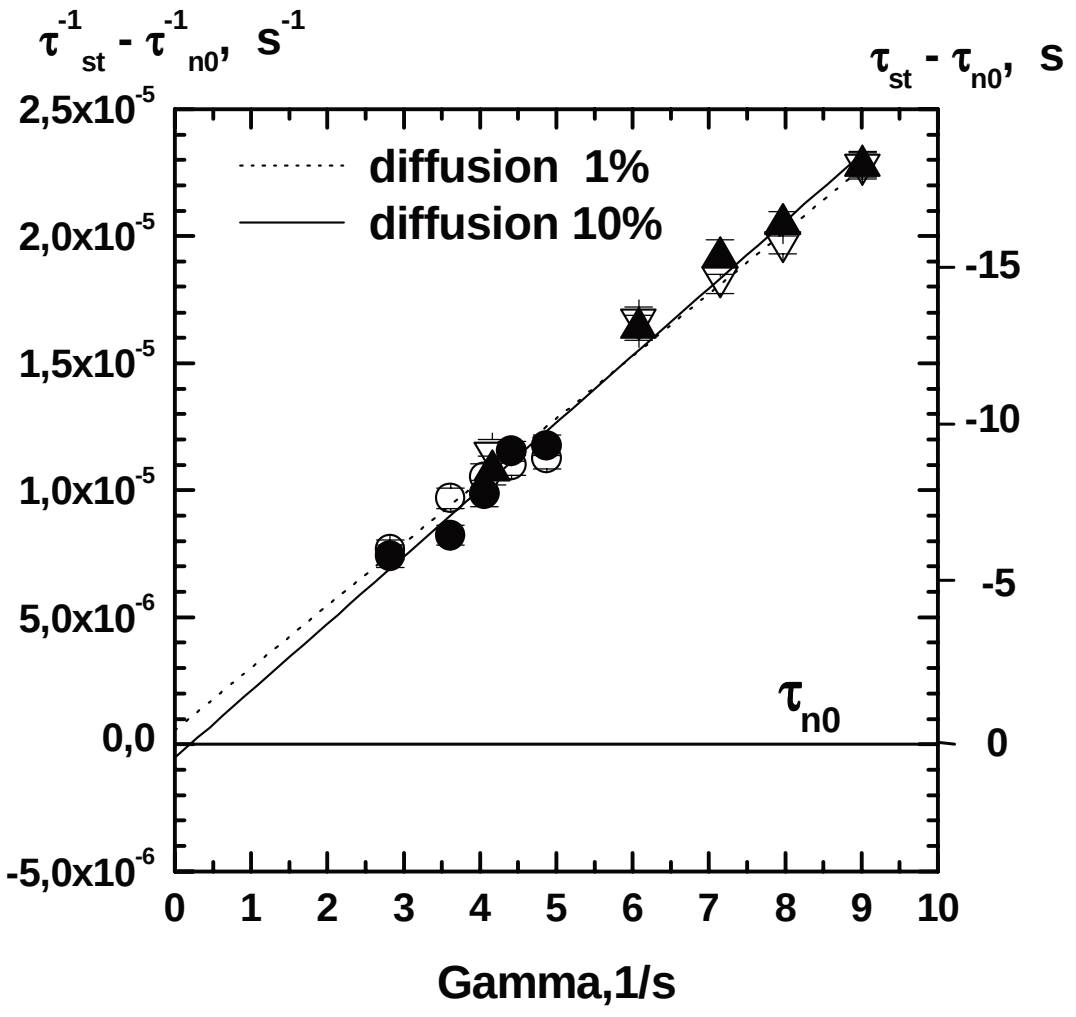

Fig. 8. Monte Carlo simulation of the experiment. Modelling the extrapolation to n-lifetime.

## 7. The influence of the residual gas for UCN storage

On the level of accuracy for neutron lifetime measurements of about 1 s, the influence of residual gas at a pressure of $5{\cdot}10^{-6}$ mbar is already important. This correction cannot be measured directly, for instance by improving the vacuum by one order of magnitude, because the expected effect is less than the statistical uncertainty. Instead, we increased the residual gas pressure to $8{\cdot}10^{-4}$ mbar, making the $(p\tau)$ – parameter for residual gas 9.5 mbar·s and obtained a calculated correction to the storage time of $0.4 \pm 0.02$ s. It does not depend on UCN energy, so this correction can be used for the neutron lifetime result.

## 8. Final result for the neutron lifetime and the list of systematic corrections and uncertainties

Values of systematic effects with their uncertainties are shown in Table 1. The main contribution to uncertainty we have is due to measurement statistics. The next largest is the uncertainty in the calculation of the function γ. The contributions due to influence of the shape of the $\mu(E)$-function and uncertainty of UCN spectrum are considerably less; these were estimated by means of variation of their parameters within the uncertainty allowed by the experimental data. Thus the total systematic correction is $0.4 \pm 0.3$ s and the final result for the neutron lifetime from our experiment is $878.5 \pm 0.7_{stat.} \pm 0.3_{syst}$.

Table 1. The systematic effects and their uncertainties

| Systematic effect | Value, s | Uncertainty, s |
|---|---|---|
| Method of $\gamma$ values calculation | 0 | 0.236 |
| Influence of $\mu(E)$ function shape | 0 | 0.144 |
| Spectrum uncertainties | 0 | 0.104 |
| Uncertainties of traps sizes(1mm) | 0 | 0.058 |
| Influence of the residual gas | 0.4 | 0.024 |
| Uncertainty of LTF critical energy (20 neV) | 0 | 0.004 |
| **Total systematic effect** | **0.4** | **0.3** |

## 9. Conclusion

In the present experiment the storage time is very close to the neutron lifetime. The difference between the best-measured storage time and the neutron lifetime is about 5s only. This gives confidence in the reliability of the result obtained.

It should be mentioned that in the experiment [1], which mainly determines the world average value, the storage times for two different trap configurations were 780 s and 500 s, i.e. the extrapolated value was 105 s. The systematic error of extrapolation to neutron lifetime was estimated by 0.4 s, i.e. 0.4% of the extrapolated difference. So high accuracy of extrapolation using two points requires special justification.

It should be mentioned also that our new result differs by 2.9 standard deviations from the result of our old experiment with solid oxygen coating, where accuracy of measurements was 4 times worse [10]. New oil coating gives more guaranties in identity of coating for wide and narrow traps than solid oxygen. (It is clear from experiments with *Ti* sublayer.) If this deviation was not random we can not exclude the case that narrow trap was coated by solid oxygen a little bit worse because of its shape is more complicated for coating than spherical one. Therefore we favour the new result of neutron lifetime measurement with LTF coating not only due to high statistical accuracy.

The new result for the neutron lifetime (878.5 ± 0.8 s) can be used for the unitarity test of Cabibbo-Kobayashi-Maskawa (CKM) matrix. Fig. 4 shows a plot of $V_{ud}$ versus $-G_A/G_V$ from [3] with the new result for the neutron lifetime. The authors of work [3] favour result $A_0$=-0.1189(7) [2] for $\beta-$asymmetry in comparison with the world average value $A_0$=-0.1173(13) [9]. (In earlier experiments the large corrections had to be made for neutron polarization, electron magnetic mirror effects and background, which were all in the 15% to 30% range.) We follow recommendations of work [3] but the $G_A/G_V$-value from the world average value $A_0$ is shown in Fig. 4 also.

The new lifetime result is different from the world average value (885.7 ± 0.8 s [9]) by 6.5 standard deviations, and by 5.6 standard deviations from the previous most precise result [1]. However, the new result for the neutron lifetime together with the current value for the $\beta-$asymmetry in neutron decay ($A_0$=-0.1189(7) [2]) is in a good agreement with the Standard Model.

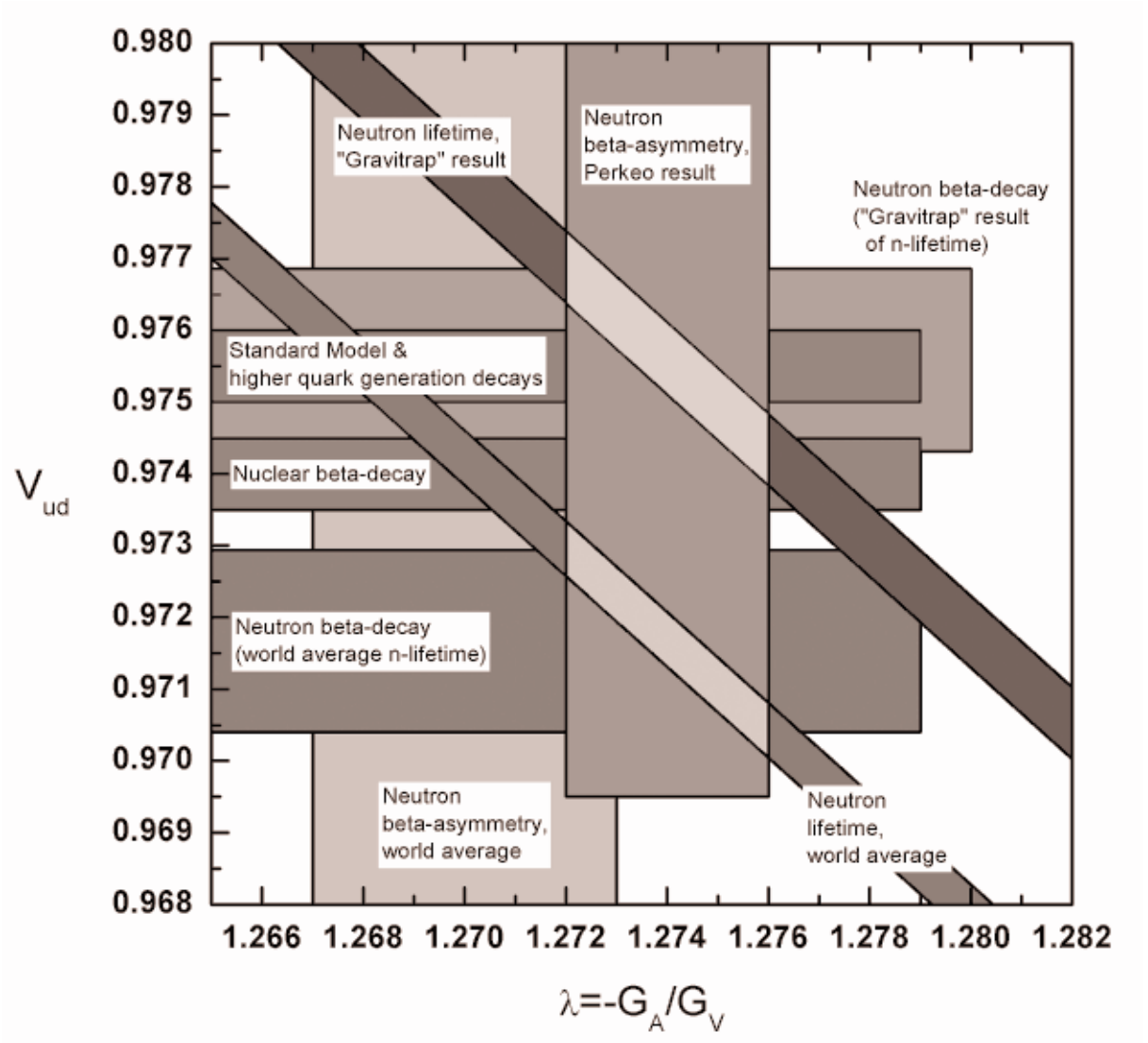

Fig. 9. $|V_{ud}|$ versus $-G_A/G_V$. $|V_{ud}|$ was derived from higher quark generation decays via $|V_{ud}|=\sqrt{1-|V_{us}|^2-|V_{ub}|^2}$ predicted from unitarity, from Ft values of nuclear-decays, and neutron β-decay.

## 10. Acknowledgements

The authors are grateful to:

- K.Schreckenbach for an opportunity to use MAMBO II position of UCN turbine;
- V.Alfimenkov, V.Lushchikov, A.Strelkov and V.Shvetsov for their contribution at the initial stage of the development of the installation;
- A.Steyerl, O.Kwon, N.Achiwa for participation in measurements and fruitful discussions;
- PSI for the help in manufacturing of UCN traps;

- T.Brenner for the intensive and very helpful assistance during the experiment;
- Russian Foundation of Basic Research for support under contract 04-02-17440;
- Russian Academy of Sciences for to the program "Physics of elementary particles";
- F.Atchison for the help in redaction of the article.